\documentclass[sigplan,10pt,nonacm]{acmart}

\setcopyright{none}
\renewcommand\footnotetextcopyrightpermission[1]{}

\acmConference[PPoPP '27]{32nd ACM SIGPLAN Annual Symposium on Principles and Practice of Parallel Programming}{March 20--24, 2027}{Salt Lake City, UT, USA}
\acmYear{2027}
\copyrightyear{2027}
\acmDOI{}
\acmISBN{}

\AtBeginDocument{%
  }

\newcommand{\textbold}[1]{\textbf{#1}}

\usepackage{enumitem}
\usepackage{booktabs}
\usepackage{makecell}
\usepackage{multirow}
\usepackage{placeins}
\usepackage{subcaption}
\usepackage{xcolor}
\usepackage{colortbl}
\definecolor{evalhead}{HTML}{D9E8F0}
\definecolor{evalbody}{HTML}{EFF5F8}

\usepackage[ruled,vlined,linesnumbered]{algorithm2e}
\newcommand{\libraauthorname}[2]{%
  \makebox[0.31\textwidth][c]{\fontsize{11}{13}\selectfont #1\textsuperscript{#2}}%
}
\newcommand{\libraauthoremail}[1]{%
  {\fontsize{9}{11}\selectfont\href{mailto:#1}{#1}}%
}
\makeatletter
\renewcommand{\@mkauthors@iii}{%
  \global\setbox\mktitle@bx=\vbox{%
    \unvbox\mktitle@bx\par\medskip
    \noindent\centering
    \setlength{\tabcolsep}{0pt}%
    \begin{tabular*}{\textwidth}{@{\extracolsep{\fill}}ccc@{}}
\libraauthorname{Shengkai Lyu}{1,2} & \libraauthorname{Hairui Zhao}{1,*} & \libraauthorname{Daran Sun}{1,2} \\[1pt]
\libraauthoremail{lvshengkai23@mails.ucas.ac.cn} & \libraauthoremail{zhaohairui@ict.ac.cn} & \libraauthoremail{sundaran24s@ict.ac.cn} \\[9pt]
\libraauthorname{Yida Gu}{1,2} & \libraauthorname{Xingchen Liu}{1,2} & \libraauthorname{YueYuan Zhou}{1} \\[1pt]
\libraauthoremail{guyida@ict.ac.cn} & \libraauthoremail{liuxingchen232@mails.ucas.ac.cn} & \libraauthoremail{zhouyueyuan@ict.ac.cn} \\[9pt]
\libraauthorname{Zheng Wei}{1} & \libraauthorname{Guangming Tan}{1} & \libraauthorname{Dingwen Tao}{1,*} \\[1pt]
\libraauthoremail{weizheng@ncic.ac.cn} & \libraauthoremail{tgm@ict.ac.cn} & \libraauthoremail{taodingwen@ict.ac.cn}
    \end{tabular*}\par
    \vspace{9pt}%
    {\small\textsuperscript{1}Institute of Computing Technology, Chinese Academy of Sciences, Beijing, China\par
     \textsuperscript{2}University of Chinese Academy of Sciences, Beijing, China\par
     \textsuperscript{*}Corresponding authors: Hairui Zhao and Dingwen Tao.\par}%
    \bigskip
  }%
}
\makeatother

\begin{document}

\title[LIBRA]{LIBRA: Runtime Co-Tuning of Batch Size and Parallelism for Time-Efficient LLM Training}

\author{Shengkai Lyu}
\affiliation{
  \institution{Institute of Computing Technology, Chinese Academy of Sciences}
  \city{Beijing}
  \country{China}
}
\affiliation{
  \institution{University of Chinese Academy of Sciences}
  \city{Beijing}
  \country{China}
}
\email{lvshengkai23@mails.ucas.ac.cn}

\author{Hairui Zhao}
\affiliation{
  \institution{Institute of Computing Technology, Chinese Academy of Sciences}
  \city{Beijing}
  \country{China}
}
\email{zhaohairui@ict.ac.cn}

\author{Daran Sun}
\affiliation{
  \institution{Institute of Computing Technology, Chinese Academy of Sciences}
  \city{Beijing}
  \country{China}
}
\affiliation{
  \institution{University of Chinese Academy of Sciences}
  \city{Beijing}
  \country{China}
}
\email{sundaran24s@ict.ac.cn}

\author{Yida Gu}
\affiliation{
  \institution{Institute of Computing Technology, Chinese Academy of Sciences}
  \city{Beijing}
  \country{China}
}
\affiliation{
  \institution{University of Chinese Academy of Sciences}
  \city{Beijing}
  \country{China}
}
\email{guyida@ict.ac.cn}

\author{Xingchen Liu}
\affiliation{
  \institution{Institute of Computing Technology, Chinese Academy of Sciences}
  \city{Beijing}
  \country{China}
}
\affiliation{
  \institution{University of Chinese Academy of Sciences}
  \city{Beijing}
  \country{China}
}
\email{liuxingchen232@mails.ucas.ac.cn}

\author{YueYuan Zhou}
\affiliation{
  \institution{Institute of Computing Technology, Chinese Academy of Sciences}
  \city{Beijing}
  \country{China}
}
\email{zhouyueyuan@ict.ac.cn}

\author{Zheng Wei}
\affiliation{
  \institution{Institute of Computing Technology, Chinese Academy of Sciences}
  \city{Beijing}
  \country{China}
}
\email{weizheng@ncic.ac.cn}

\author{Guangming Tan}
\affiliation{
  \institution{Institute of Computing Technology, Chinese Academy of Sciences}
  \city{Beijing}
  \country{China}
}
\email{tgm@ict.ac.cn}

\author{Dingwen Tao}
\affiliation{
  \institution{Institute of Computing Technology, Chinese Academy of Sciences}
  \city{Beijing}
  \country{China}
}
\email{taodingwen@ict.ac.cn}

\renewcommand{\shortauthors}{Lyu et al.}

\begin{abstract}
Training large language models (LLMs) requires choosing both a batch size and a parallel configuration. Existing systems make these decisions in isolation, tuning parallelism for throughput and batch size for convergence, but neither objective alone determines training efficiency. We present LIBRA, a lightweight runtime framework that co-tunes batch size and parallel configuration to minimize end-to-end time. LIBRA treats throughput and convergence as two noisy objectives and searches their joint space using parallel noisy expected hypervolume improvement (qNEHVI). To make this runtime search efficient, we propose hierarchical pruning, which removes infeasible and low-throughput parallel configurations through system-level performance estimation and then narrows batch-size candidates through critical-batch-size-aware convergence monitoring. We further design acquisition-guided parallel evaluation (AGPE), which evaluates surviving candidates concurrently across GPU groups in order of their acquisition values. On clusters of up to 256 GPUs, LIBRA reduces time-to-convergence by 31.1\% on average over state-of-the-art methods, while runtime tuning accounts for only 0.3--3.5\% of overall time.
\end{abstract}

\maketitle

\section{Introduction}
Large language models (LLMs)~\cite{gpt3,gpt4,llama1,llama2,llama3,gemini}
have driven rapid progress across many applications. Their training increasingly
relies on large GPU clusters. For example, LLaMA~3 405B~\cite{llama3} was trained
on up to 16K H100 GPUs. Scaling such workloads requires combining data
parallelism (DP)~\cite{sergeev2018horovod}, tensor parallelism
(TP)~\cite{shoeybi2019megatron}, pipeline parallelism
(PP)~\cite{huang2019gpipe}, and context parallelism
(CP)~\cite{jacobs2023ulysses}, together with the choice of micro-batch size
(MBS) and global batch size (GBS), creating a large configuration space that
must be tuned.

\textbf{Why co-tuning is needed?} Training performance is reported using
step time, token throughput~\cite{narayanan2021efficient,zheng2022alpa},
or model FLOPs utilization (MFU)~\cite{llama3}, rather than
time-to-convergence. \textit{However, maximizing system throughput does not necessarily minimize end-to-end (E2E) time.}
Tab.~\ref{tab:e2e-compare} compares the three tuning goals.
Tuning for throughput alone achieves the highest throughput but requires more
tokens to reach the target loss. Tuning for convergence alone requires the
fewest tokens but has much lower throughput. Neither goal minimizes E2E time.
Tuning for both achieves neither the highest throughput nor the fewest tokens,
yet reaches the target loss 30.8--34.8\% sooner than tuning for throughput alone
and 44.1--50.2\% sooner than tuning for convergence alone.
\textit{Thus, E2E training depends on both processing rate (i.e., throughput)
and progress per token (i.e., convergence).}

\begin{table}[t]
    \centering
    \caption{Configurations selected under three tuning objectives: throughput,
convergence, and both (Thr.\&Conv.). All configurations reach the same target
loss for a given model, so total training time is directly comparable.
}
    \label{tab:e2e-compare}
    \footnotesize
    \setlength{\tabcolsep}{1.0pt}
    \renewcommand{\arraystretch}{1.0}
    \begin{tabular}{@{}lccrrr@{}}
        \toprule
        \makecell{\textbf{Model}\\\textbf{(target loss)}} & \textbf{GPUs} &
        \makecell{\textbf{Tuning}\\\textbf{goal}} &
        \makecell{\textbf{Throughput}\\\textbf{(K tok/s)}} &
        \makecell{\textbf{Tokens}\\\textbf{to target (B)}} &
        \makecell{\textbf{Time}\\\textbf{(h)}} \\
        \midrule
        \multirow{3}{*}{\makecell[l]{Code LLaMA 34B\\
        ($L_{\mathrm{val}}\!\rightarrow\!2.25$)}}
        & \multirow{3}{*}{128} & Throughput  & \textbf{154.16} & 70.93 & 127.80 \\
        &                      & Convergence & 58.41 & \textbf{33.28} & 158.28 \\
        &                      & Thr.\&Conv. & 121.02 & 38.55 & \textbf{88.48} \\
        \midrule
        \multirow{3}{*}{\makecell[l]{LLaMA-2 70B\\
        ($L_{\mathrm{val}}\!\rightarrow\!2.22$)}}
        & \multirow{3}{*}{256} & Throughput  & \textbf{82.35} & 88.76 & 299.42 \\
        &                      & Convergence & 24.66 & \textbf{34.82} & 392.28 \\
        &                      & Thr.\&Conv. & 58.8 & 41.32 & \textbf{195.22} \\
        \bottomrule
    \end{tabular}
\end{table}

However, throughput and convergence are coupled through the training
configuration and jointly determine E2E training efficiency.
Specifically, each DP replica executes $\mathrm{GBS}/(\mathrm{DP}\times\mathrm{MBS})$ micro-batches, so a
small GBS limits DP and leaves little work to amortize PP bubbles. MBS
controls per-GPU utilization. Keeping pipelines busy commonly requires
$\mathrm{GBS}\geq2\times\mathrm{PP}\times\mathrm{DP}\times\mathrm{MBS}$,
potentially exceeding the convergence-efficient GBS. For LLaMA-2 13B, constant
GBS delivers 112.00K tokens/s but consumes 51.32B tokens. Ramp-up
($32\rightarrow128\rightarrow512$) reduces the token demand to 34.00B,
parallel tuning raises its throughput from 83.13K to 100.46K tokens/s, and
joint co-tuning reaches the target with 28.27B tokens in 81.37 hours. Thus,
batch size must adapt
jointly with the parallel configuration during training.

\textbf{Limitations of existing methods.} Parallel auto-tuners
such as Alpa~\cite{zheng2022alpa},\allowbreak{} Aceso~\cite{aceso},\allowbreak{}
FlexFlow~\cite{flexflow},\allowbreak{} and Hypertron~\cite{hypertron} optimize
execution throughput but exclude batch size and its convergence effect. Conversely, multi-fidelity hyperparameter
optimization (HPO) methods
such as Hyperband~\cite{li2017hyperband},\allowbreak{}
ASHA~\cite{li2018massively},\allowbreak{} BOHB~\cite{falkner2018bohb},
and DEHB~\cite{awad2021dehb} prune unpromising batch-size trials
while fixing parallel configurations.
In practice,
LLaMA~3~\cite{llama3}, DeepSeek-V3~\cite{deepseek-v3}, and
MiniMax-01~\cite{minimax-01} use model-specific handcrafted ramp-up schedules,
while branched training~\cite{merrill2025critical} estimates critical batch
size (CBS) from partial runs. Neither strategy transfers across models nor
adapts during training. Pollux~\cite{qiao2021polluxcoadaptiveclusterscheduling}
comes closest by co-adapting batch size and resource allocation, yet it targets
cluster scheduling and does not search parallel configurations. In
short, no existing approach cheaply co-tunes batch size and parallel
configuration online.

\textbf{Research challenges.} These limitations call for a lightweight runtime
mechanism that co-tunes batch size and parallel configuration as training
evolves. This presents two challenges. First, \textit{the joint configuration
space is large and evolves during training}. It contains approximately 1.5M
candidates on 64 GPUs and 1.9M on 128 GPUs, while feasibility checks still
leave more than 11K candidates. Second, \textit{runtime evaluation is
expensive}. Each candidate occupies multiple GPUs and must be observed for
throughput and convergence, making sequential evaluation impractical.

\textbf{Our solution.} To address these challenges, we present LIBRA, a
lightweight runtime framework that co-tunes batch size and parallel
configuration to optimize throughput and convergence. LIBRA models both
as noisy objectives and uses parallel noisy expected hypervolume improvement
(qNEHVI)~\cite{daulton2021qnehvi} to select promising configurations. To tame
the large and shifting search space (challenge 1), hierarchical pruning
combines performance estimation and throughput-aware pruning with convergence
monitoring and CBS-aware GBS pruning. To keep runtime evaluation affordable (challenge 2),
acquisition-guided parallel evaluation (AGPE) evaluates retained candidates
concurrently across GPU groups and feeds asynchronous observations to the
optimizer. LIBRA then applies the configuration during training.
This paper makes the following contributions:

\begin{itemize}[leftmargin=*,topsep=3pt]
\item We identify joint throughput--convergence
tuning as a missing layer in large-scale LLM training, and formulate the
selection of batch size and parallel configuration as a noisy multi-objective
optimization problem, which we solve with qNEHVI-based Bayesian optimization.

\item We make runtime co-tuning
lightweight by exploiting the fact that the useful region of the search
space shifts as training progresses: LIBRA prunes
low-performance parallel configurations via performance estimation, then
narrows batch-size candidates online through convergence monitoring and
CBS-aware GBS pruning.

\item We design acquisition-guided parallel evaluation (AGPE), which
selects and places candidates using qNEHVI acquisition values, GPU demand, and
estimated evaluation time, and returns observations to the optimizer
asynchronously.

\item  On a range of LLMs at up to 256 GPUs, LIBRA reduces
end-to-end time-to-convergence by 31.1\% on average over state-of-the-art
methods, with runtime tuning costing only 0.3--3.5\% of end-to-end training
time.
\end{itemize}
\section{Background}

\subsection{Throughput-oriented Parallel Configurations}
Distributed LLM training combines several parallel dimensions. Data parallelism
(DP)~\cite{sergeev2018horovod} partitions batches and synchronizes gradients.
Tensor parallelism (TP)~\cite{shoeybi2019megatron} partitions intra-layer
operators. Pipeline parallelism
(PP)~\cite{huang2019gpipe,narayanan2019pipedream} divides layers into stages,
and context parallelism (CP)~\cite{liu2023ringattention,jacobs2023ulysses}
partitions long sequences. State-sharding methods, including the Zero Redundancy
Optimizer (ZeRO)~\cite{rajbhandari2020zero} and fully sharded data parallelism
(FSDP)~\cite{zhao2023fsdp}, distribute model states.

MBS controls per-GPU memory use and kernel efficiency, while GBS determines how
many micro-batches and DP replicas participate in each optimizer
step. Together, these choices determine communication, pipeline bubbles, GPU
utilization, and throughput. Systems such as Alpa~\cite{zheng2022alpa} and
Aceso~\cite{aceso} use profiling or cost models to select feasible
high-throughput configurations.

\subsection{Convergence-oriented Batch-size Ramp-up}
Larger batches reduce gradient variance and expose more data parallelism, but
eventually yield little reduction in optimizer steps. For per-example gradient
$g_i$, mean $G=\mathbb{E}[g_i]$, and covariance
$\Sigma=\operatorname{Cov}(g_i)$, McCandlish et
al.~\cite{mcandlish2018empirical} define the simplified \emph{gradient noise
scale} (GNS) as $B_{\mathrm{noise}} =
\operatorname{tr}(\Sigma)/{\lVert G\rVert_2^2}$, which approximates the
\emph{critical batch size} (CBS)~\cite{mcandlish2018empirical,shallue2019measuring}.
CBS marks where the trade-off between optimizer steps and tokens turns: below
it, enlarging the batch buys a nearly proportional reduction in steps, whereas
above it the step savings diminish while the tokens consumed keep growing. It
is thus a transition between time and token efficiency rather than a hard
convergence limit~\cite{bottou2018optimizationmethodslargescalemachine}.
Because GNS generally grows as the loss decreases, the useful batch size grows
during training. Branched training~\cite{merrill2025critical} confirms this
trend for OLMo 1B/7B, reporting slightly better OLMo-1B loss with 43\% fewer
optimizer steps under CBS-guided ramp-up.

Prior work adapts batch size at several granularities: partial-training HPO via Hyperband~\cite{li2017hyperband} and ASHA~\cite{li2018massively}, in-run rules such as AdaBatch~\cite{devarakonda2017adabatch} and coupled batch-size and learning-rate scaling~\cite{smith2018dontdecay}, and LLM-scale branched training~\cite{merrill2025critical}. In production, however, schedules are still hand-tuned per model: LLaMA~3~\cite{llama3} steps token batches 4M$\rightarrow$8M$\rightarrow$16M, DeepSeek-V3~\cite{deepseek-v3} grows the sequence batch from 3,072 to 15,360 over 469B tokens, and MiniMax-01~\cite{minimax-01} uses 16M$\rightarrow$32M$\rightarrow$64M$\rightarrow$128M.

\subsection{Multi-objective Optimization Methods}
Multi-objective optimization balances conflicting objectives. Pareto-optimal
configurations trade off objectives, and their objective vectors form the
Pareto frontier~\cite{miettinen1999nonlinear}.

\textbold{Scalarization Methods.} Weighted-sum and Chebyshev utilities reduce
multiple objectives to one~\cite{miettinen1999nonlinear}, but depend on
normalization and fixed preferences. Weighted sums may miss unsupported points
on non-convex frontiers.

\textbold{Pareto-based Evolutionary Methods.} NSGA-II~\cite{deb2002nsga2} uses
non-dominated sorting and crowding distance, while SPEA2~\cite{zitzler2001spea2}
uses strength-based fitness, density, and an archive. Such population methods
support discrete, non-convex spaces but usually require many evaluations.

\textbold{Surrogate-based Multi-objective Bayesian Optimization.}
Multi-objective Bayesian optimization uses probabilistic surrogates and
acquisition functions for costly observations. ParEGO~\cite{knowles2006parego}
applies expected improvement to randomized Chebyshev scalarizations. Expected
hypervolume improvement (EHVI)~\cite{emmerich2011ehvi} measures expected
hypervolume growth. qNEHVI~\cite{daulton2021qnehvi} extends EHVI to noisy,
batched selection by integrating uncertainty in the latent frontier.

\begin{figure}[tbp]
    \centering
    \includegraphics[width=\columnwidth]{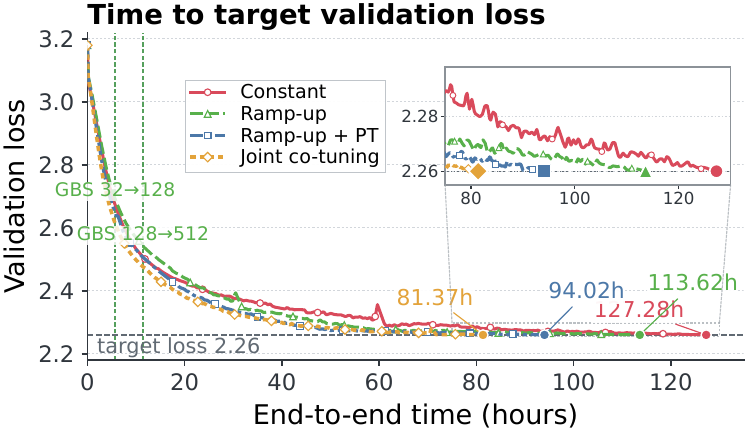}
    \caption{Comparison of constant GBS, GBS ramp-up, GBS ramp-up with parallel
    tuning (PT), and joint co-tuning for LLaMA-2 13B. The curves show validation
    loss over E2E time and mark GBS changes and time to target.}
    \label{fig:motivation-rampup}
    \Description{Validation loss over time for four policies. Joint co-tuning
    reaches the target first by selecting GBS and parallel configuration
    together.}
\end{figure}

\section{Motivation and Challenges}
\label{sec:motivation_challenges}
This section shows why throughput and convergence must be optimized jointly
during training, then summarizes the limitations and challenges that motivate
LIBRA.

\subsection{Why Co-Tune Throughput and Convergence?}

\textbf{Throughput alone does not minimize time-to-\linebreak
convergence.}
Tab.~\ref{tab:e2e-compare} compares three tuning goals for Code LLaMA 34B and
LLaMA-2 70B. Tuning for throughput attains the highest processing rate, and
tuning for convergence requires the fewest tokens, yet both finish later than
tuning for the two jointly, which reaches the same target loss 30.8--34.8\%
sooner than throughput-only tuning. \space
Time to the target is the number of tokens required divided by average token
throughput. GBS affects both quantities. A lower-throughput configuration can thus finish earlier by
requiring fewer tokens, so throughput and convergence must be retained as
separate objectives throughout the configuration search.

\textbf{Batch size and parallel configuration are coupled, and their coupling
shifts during training.}
Since $\mathrm{GBS} = \mathrm{DP} \times \mathrm{MBS} \times$ (accumulated
micro-batches), a small GBS caps DP and may leave too few micro-batches to
amortize PP bubbles, while a larger GBS exposes more data parallelism;
conversely, TP, PP, CP, and MBS fix the memory, communication, and pipeline cost
of any given GBS. For LLaMA-2 13B on 64 GPUs, raising GBS from 16 to 256 shifts
the selected DP from 2 to 32 and the active GPU allocation from 8 to 64 GPUs, so the
throughput-optimal parallel configuration is a function of the batch decision.
This dependence also evolves over training (Fig.~\ref{fig:motivation-rampup}):
constant GBS attains the highest throughput yet needs 51.32B tokens and 127.28
hours; ramping GBS $32\!\rightarrow\!128\!\rightarrow\!512$ cuts token demand and
reaches the target in 113.62 hours, and retuning the parallel configuration along
the same ramp-up lowers this to 94.02 hours. Because that combination still
chooses the two separately, it is surpassed by joint co-tuning, which finishes in
81.37 hours at a \emph{lower} average throughput (96.50K vs.\ 100.46K tokens/s)
by cutting token demand to 28.27B---confirming that GBS and parallel
configuration must be selected together at runtime.

\subsection{Limitations of Existing Tuning Paradigms}
Existing methods cover \allowbreak\space
different parts of the joint search.
Tab.~\ref{tab:existing-work} \space
compares which decisions they update during training.

\textbold{Parallel Configuration Search.}
Parallel configuration search selects DP, TP, PP, CP, and MBS together.
Systems such as Alpa~\cite{zheng2022alpa}, Aceso~\cite{aceso},
FlexFlow~\cite{flexflow}, and Hypertron~\cite{hypertron} search parallel
configurations to reduce iteration time or increase throughput. Their search
assumes a specified computation and leaves GBS and its effect on convergence
outside the optimization. Given an external batch schedule, these systems can
find a fast parallel configuration for each prescribed GBS, but cannot determine
when to change GBS or whether that choice minimizes time-to-convergence.

\begin{table}[tbp]
    \centering
    \caption{Comparison of existing methods and LIBRA.}
    \label{tab:existing-work}
    \scriptsize
    \setlength{\tabcolsep}{1.6pt}
    \renewcommand{\arraystretch}{1.02}
    \begin{tabular}{@{}ccccc@{}}
        \toprule
        \textbf{Method} &
        \makecell{\textbf{Parallelism}\\\textbf{Search}} &
        \makecell{\textbf{Convergence}\\\textbf{Optimization}} &
        \makecell{\textbf{GBS-Parallel}\\\textbf{Co-tuning}} &
        \makecell{\textbf{Online}\\\textbf{Tuning}} \\
        \midrule
        Alpa~\cite{zheng2022alpa} & \ding{51} & \ding{55} & \ding{55} & \ding{55} \\
        Aceso~\cite{aceso} & \ding{51} & \ding{55} & \ding{55} & \ding{55} \\
        ASHA~\cite{li2018massively} & \ding{55} & \ding{51} & \ding{55} & \ding{51} \\
        BOHB~\cite{falkner2018bohb} & \ding{55} & \ding{51} & \ding{55} & \ding{51} \\
        Handcrafted ramp-up~\cite{llama3} & \ding{55} & \ding{51} & \ding{55} & \ding{55} \\
        Branched training~\cite{merrill2025critical} & \ding{55} & \ding{51} & \ding{55} & \ding{55} \\
        Pollux~\cite{qiao2021polluxcoadaptiveclusterscheduling} & \ding{55} & \ding{51} & \ding{55} & \ding{51} \\
        \midrule
        \textbf{LIBRA (ours)} & \ding{51} & \ding{51} & \ding{51} & \ding{51} \\
        \bottomrule
    \end{tabular}
\end{table}

\textbold{Convergence Optimization and Online Tuning.}
Convergence optimization considers training progress when selecting
configurations. Online tuning uses completed evaluations to update later tuning
decisions. Multi-fidelity methods such as Hyperband~\cite{li2017hyperband},
ASHA~\cite{li2018massively}, BOHB~\cite{falkner2018bohb}, and
DEHB~\cite{awad2021dehb} allocate partial-training resources across
hyperparameter configurations and stop weak runs early. ASHA and BOHB use
partial-run observations to update later tuning decisions, but keep the
parallel configuration fixed. Applying these methods to parallel-configuration
search introduces costly cluster-level evaluations because each candidate
occupies multiple GPUs and must run long enough to expose convergence
differences. Early stopping reduces the duration of runs but does not avoid
these evaluations.

Pollux~\cite{qiao2021polluxcoadaptiveclusterscheduling} adapts batch size and
resource allocation online using throughput and statistical efficiency.
However, it neither searches TP/PP/CP/MBS configurations nor preserves
throughput and convergence as separate objectives.

\begin{figure}[tbp]
    \centering
    \includegraphics[width=\columnwidth]{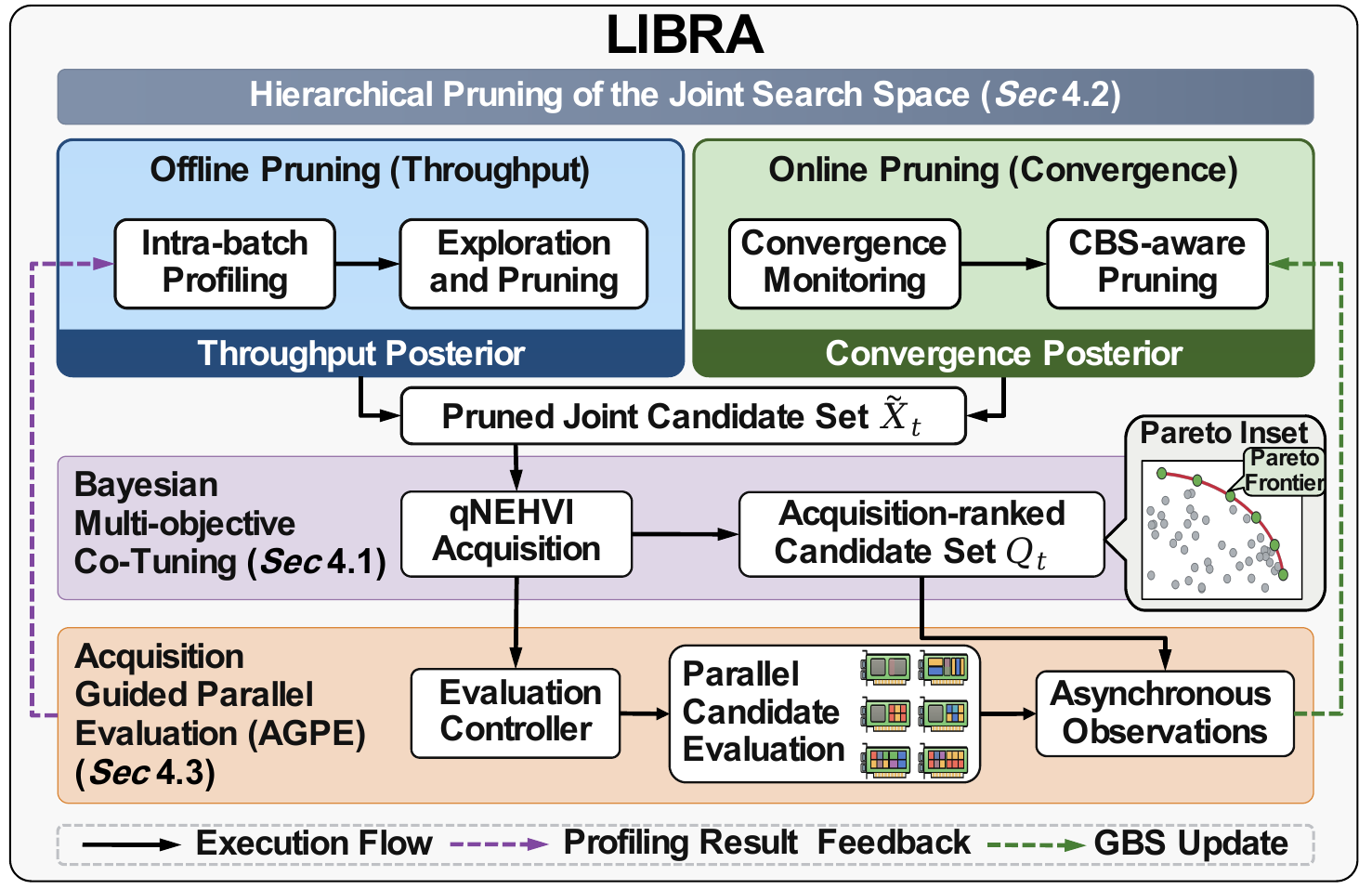}
    \caption{Overview of LIBRA's key design.}
    \label{fig:libra_overview}
    \Description{Overview of LIBRA's hierarchical pruning, Bayesian co-tuning,
    and acquisition-guided parallel evaluation workflow.}
\end{figure}

\textbold{Handcrafted and Branched Batch Tuning.}
GBS-parallel co-tuning selects GBS and parallel configuration together.
Production LLM runs commonly prescribe batch-size schedules for a particular
training recipe, as illustrated by LLaMA~3~\cite{llama3},
DeepSeek-V3~\cite{deepseek-v3}, and MiniMax-01~\cite{minimax-01}. These
schedules select batch-size transitions for one model and recipe, while leaving
the parallel configuration unchanged or re-tuning it separately after each
batch change. Branched training~\cite{merrill2025critical} estimates CBS from
multiple partial runs, but replicating branches from shared checkpoints remains
expensive at cluster scale. These approaches use a fixed schedule or separate
branches rather than completed evaluations from the ongoing training run, and
neither provides GBS-parallel co-tuning.

\textit{In summary, high training efficiency requires a lightweight runtime
tuner that co-optimizes throughput and convergence.}

\subsection{Research Challenges}

To realize this design, LIBRA must reduce a large joint domain and obtain costly
runtime observations efficiently.

\textbold{Large Joint Search Space of Parallelism and Batch Size.}
The cross product of GBS, MBS, DP, TP, PP, and CP is constrained but
combinatorial. Under the discretization used by LIBRA, the space contains
approximately 1.5M configurations on 64 GPUs and 1.9M on 128 GPUs. Across a
full parameter sweep of models and allowed ranges, feasibility filtering leaves
11K--360K candidates at 64 GPUs and 15K--480K at 128 GPUs. These counts describe
the broad sweep. The evaluated domains use narrower model-specific ranges and
contain 11.3K--24.6K executable candidates. Moreover, system measurements can remove infeasible or slow
configurations, but throughput cannot safely eliminate a batch whose convergence benefit
may change later. Complementary system- and convergence-side pruning must
therefore preserve uncertainty for the Bayesian optimizer.

\textbold{Inefficiency of Sequential Configuration Evaluation.}
Evaluating a candidate requires measured system behavior and a bounded training
window for convergence progress. Evaluations use different GPU counts and
durations. Sequential execution adds every cost to the critical path, whereas
a fixed number of concurrent evaluations can fragment GPUs or run low-value candidates. Runtime
tuning must prioritize informative candidates, pack candidates by GPU demand and estimated evaluation time, and return observations asynchronously.
Section~\ref{sec:joint-optimization} addresses the first challenge with hierarchical
pruning and the second with acquisition-guided parallel evaluation.

\section{Iterative Joint Optimization for Throughput and Convergence}
\label{sec:joint-optimization}
LIBRA preserves throughput and convergence as separate objectives,
hierarchically prunes the configuration space, and then evaluates candidates
through acquisition-guided parallel evaluation (AGPE).
Fig.~\ref{fig:libra_overview} shows the overview of LIBRA's key design.

\subsection{Problem Formulation and Bayesian Optimization}
\label{sec:bayesian-cotuning}
\textbold{Problem Formulation.}
Let $s_t$ denote the training state and $x_t$ the configuration selected at
decision point $t$. LIBRA minimizes the time to target validation loss
$\ell^{\star}$ as follows:
\begin{equation}
\small
    T^{\star}
    =
    \min_{\{x_t\}}
    \inf\{\tau \geq 0\mid\ell_{\mathrm{val}}(s(\tau))\leq\ell^{\star}\},
    \quad x_t\in\mathcal{X}(s_t),
    \label{eq:time-to-convergence-objective}
\end{equation}
where $s(\tau)$ is the induced training trajectory and $\mathcal{X}(s_t)$ is the
feasible set. Solving Eq.~\ref{eq:time-to-convergence-objective} over
configuration sequences requires many branched runs. LIBRA instead makes local
decisions from token throughput $f_{\mathrm{thr}}$ and convergence progress per
token $f_{\mathrm{conv}}$ at the current state.

A candidate is $x=(b,\pi)$, where $b$ is GBS and
$\pi=(DP,TP,PP,CP,MBS)$ is its parallel configuration. Candidates satisfy
device, memory, divisibility, and execution constraints. Each starts from a
common checkpoint $c_t$ and runs for $N_x$ tokens. LIBRA measures validation
loss on the same validation data before and after monitoring, yielding
\begin{equation}
\small
    \widehat{f}_{\mathrm{conv}}(x\mid s_t)
    =
    \frac{\ell(c_t)-\ell(c_t^x)}{N_x},
    \label{eq:convergence-progress}
\end{equation}
{}\space where $c_t^x$ is the checkpoint reached by candidate $x$. Each optimizer
step processes $bN_{\mathrm{tok}}$ tokens, where $N_{\mathrm{tok}}$ is the
sequence length. At a fixed $b$, changing the layout changes the time for
that step while preserving the examples in the update. LIBRA therefore
measures convergence once per GBS at $c_t$. A later checkpoint supplies a
new observation as the model learns. A shared checkpoint and validation data make
observations comparable, while runtime variation and short horizons make both
objectives noisy.

\textbold{Bayesian Multi-objective Optimization.}
LIBRA applies multi-objective Bayesian optimization over the finite set
$\mathcal{X}_t$. Completed evaluations form
$\mathcal{D}_t=\{(x_i,\mathbf{y}_i,\Sigma_i)\}_{i=1}^{n_t}$, where
$\mathbf{y}_i$ contains both objectives and $\Sigma_i$ their observation-noise
covariance. LIBRA maintains separate Gaussian-process posteriors for throughput
and convergence. Performance estimates initialize the former. Runtime monitoring
updates the latter. Goodput, the product of throughput and statistical
efficiency~\cite{qiao2021polluxcoadaptiveclusterscheduling}, provides a fixed
scalarization, while Pareto-based evolutionary methods such as
NSGA-II~\cite{deb2002nsga2} and SPEA2~\cite{zitzler2001spea2} require many
evaluations. Among Bayesian methods,
ParEGO~\cite{knowles2006parego} still optimizes sampled scalarizations, and
standard EHVI~\cite{emmerich2011ehvi} does not account for noisy observations or
parallel selection. Parallel noisy expected hypervolume improvement
(qNEHVI)~\cite{daulton2021qnehvi} handles both, matching LIBRA's costly, noisy,
and concurrent runtime evaluations.

The non-dominated observations form a Pareto frontier $\mathcal{P}_t$. LIBRA
uses qNEHVI to select candidates expected to extend this frontier while
accounting for observation noise and posterior uncertainty. For a batch $X$ of
$q$ candidates,
\begin{equation}
\small
\begin{split}
    \alpha_{q\mathrm{NEHVI}}(X)
    =
    \mathbb{E}_{\mathbf{f}_t\mid\mathcal{D}_t}
    \big[&\mathrm{HV}_{\mathbf{r}}(\mathcal{P}_t(\mathbf{f}_t)
       \cup\mathbf{f}_t(X)) \\
       &-\mathrm{HV}_{\mathbf{r}}(\mathcal{P}_t(\mathbf{f}_t))\big],
    \label{eq:qnehvi}
\end{split}
\end{equation}
where $\mathbf{f}_t$ is a posterior function sample,
$\mathcal{P}_t(\mathbf{f}_t)$ its non-dominated frontier, and
$\mathrm{HV}_{\mathbf{r}}(\cdot)$ the hypervolume relative to reference point
$\mathbf{r}$. LIBRA estimates Eq.~\ref{eq:qnehvi} with Monte
Carlo samples and greedily batches candidates from the pruned set
$\widetilde{\mathcal{X}}_t$ using marginal qNEHVI values. Running evaluations
remain pending to avoid resampling the same region.

qNEHVI ranks evaluations rather than choosing the deployed configuration. Let
$\Delta\ell_t$ be the fixed loss decrement from the current validation loss to
the next loss milestone, shared by all candidates at $s_t$. For posterior sample
$j$, LIBRA estimates the local time of an evaluated candidate as
\begin{equation}
\small
    \widehat{T}^{(j)}_t(x)
    =t_{\mathrm{reconf}}(x)
    +\frac{\Delta\ell_t}
    {f^{(j)}_{\mathrm{thr}}(x)f^{(j)}_{\mathrm{conv}}(x)},
    \label{eq:deployment-time}
\end{equation}
where $t_{\mathrm{reconf}}(x)$ is checkpoint and reconfiguration time and is
zero for $x_t$. If $f^{(j)}_{\mathrm{conv}}(x)\leq0$, LIBRA sets
$\widehat{T}^{(j)}_t(x)=+\infty$, so that sample cannot support a switch. For an
evaluated non-dominated candidate, each sample gives the time reduction
$\Delta T^{(j)}_t(x)=\widehat{T}^{(j)}_t(x_t)-\widehat{T}^{(j)}_t(x)$. The
controller switches only when the fifth posterior percentile, a 95\% lower
bound, of $\Delta T_t(x)$ is positive. Otherwise, it keeps $x_t$. \space
The product in Eq.~\ref{eq:deployment-time} gives the expected loss decrease
per second at the candidate's deployment throughput. A larger GBS can take
longer to reach the next milestone even when it processes more tokens per
second. At the next decision point, LIBRA updates the estimate from the
selected training state.

\subsection{Hierarchical Pruning of the Joint Search Space}
\label{sec:hierarchical-pruning}
\begin{figure}[t]
    \centering
    \includegraphics[width=.95\columnwidth]{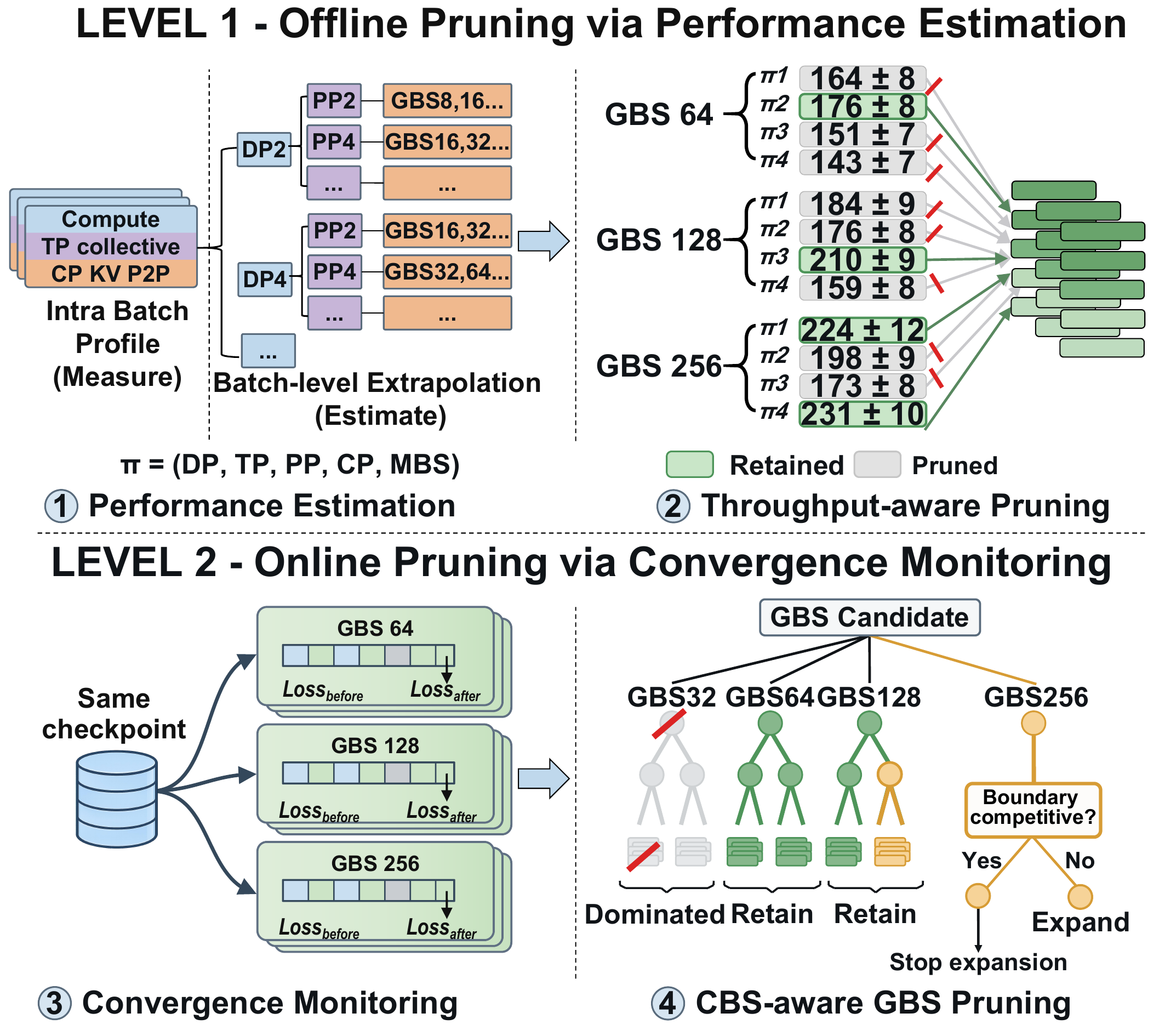}
    \caption{Overview of hierarchical pruning in LIBRA.
The first level ranks layouts for each GBS and GPU count. The second follows
the progress of distinct batch sizes from a common checkpoint.}
    \label{fig:throughput-extrapolation}
    \Description{A four-stage diagram of performance estimation from sparse
    intra-batch measurements, throughput ranking and pruning, convergence
    monitoring from a shared checkpoint, and CBS-aware GBS pruning.}
\end{figure}

Applying qNEHVI to the cross product of GBS and parallel configurations is
impractical. LIBRA therefore uses the two pruning levels in
Fig.~\ref{fig:throughput-extrapolation}. Level 1 estimates performance and
prunes parallel configurations within each GBS. Level 2 monitors convergence
and prunes the ordered GBS pool. They retain uncertain candidates for qNEHVI
selection.

\textbf{Offline Pruning via Performance Estimation.}
LIBRA enumerates candidates satisfying
$DP\cdot TP\cdot PP\cdot CP\le W$, where $W$ is the GPU count, and
requires $b$ to be divisible by $DP\cdot MBS$. Before profiling, a memory bound
with a safety margin removes configurations prone to out-of-memory (OOM)
failures. Let $N_{\mathrm{stage}}(PP)$ and $L_{\mathrm{stage}}(PP)$ be the
largest parameter and layer counts per pipeline stage, $S$ and $H$
the sequence length and hidden size, and
$b_p$, $b_g$, $b_o$, $b_m$, and $b_a$ the storage costs of parameters,
gradients, optimizer states, master weights, and activations. LIBRA estimates
\begin{equation}
\small
\begin{split}
\widehat{M}_{\mathrm{peak}}(x)
={}&\frac{N_{\mathrm{stage}}(PP)}{TP}
\left(b_p+\frac{b_g}{z_g(x)}+\frac{b_o}{z_o(x)}
+\frac{b_m}{z_m(x)}\right) \\
&+MBS\cdot\frac{S}{CP}\cdot H\cdot L_{\mathrm{stage}}(PP)
  \cdot\frac{\kappa(TP,CP)}{TP}\cdot b_a,
\end{split}
\label{eq:memory-bound}
\end{equation}
where $z_g(x)$, $z_o(x)$, and $z_m(x)$ are DP sharding factors for gradients,
optimizer states, and master weights. Each is 1 when replicated and $DP$ when
sharded. $\kappa(TP,CP)$ captures activation checkpointing, sequence
parallelism, and temporary buffers. Using Eq.~\ref{eq:memory-bound}, LIBRA \space
directly admits layouts for which
$\widehat{M}_{\mathrm{peak}}(x)\le0.9M_{\mathrm{GPU}}$, where
$M_{\mathrm{GPU}}$ is memory capacity. Across 451 launches, configurations remained OOM-safe.
Near device capacity, temporary buffers and the number of micro-batches in
flight can determine whether a layout fits. LIBRA can resolve these cases
with complete optimizer steps at the proposed MBS, filling the pipeline and
recording peak memory on every rank. A layout that completes the steps
enters the throughput comparison.

\textit{Intra-batch Profiling and Batch-level Extrapolation.}
Module 1 in Fig.~\ref{fig:throughput-extrapolation} measures
$(TP,CP,MBS)$ configurations. Profiles record micro-batch compute, TP
collective, and CP key-value (KV) exchange latency on the smallest feasible GPU
group. LIBRA repeats each profile 3--5 times, discards the warm-up step, and
estimates mean and variance. Candidates sharing the $(TP,CP,MBS)$ tuple reuse
this profile.

Tab.~\ref{tab:sparse-profiling} shows that sparse profiling measures 68--98
representatives, 0.40--0.60\% of the feasible search space.

\begin{table}[t]
    \centering
    \caption{Comparison of the feasible search space and measured profiles
    across workloads.}
    \label{tab:sparse-profiling}
    \footnotesize
    \setlength{\tabcolsep}{4.0pt}
    \renewcommand{\arraystretch}{1.0}
    \begin{tabular}{@{}lccr@{}}
        \toprule
        \textbf{Workload} & \textbf{Search space} &
        \textbf{Measured profiles} & \textbf{Coverage} \\
        \midrule
        7B@64 GPUs   & 19,872 & 84 & 0.42\% \\
        13B@64 GPUs  & 11,264 & 68 & 0.60\% \\
        34B@128 GPUs & 15,360 & 92 & 0.60\% \\
        70B@256 GPUs & 24,576 & 98 & 0.40\% \\
        \bottomrule
    \end{tabular}

\end{table}

LIBRA extrapolates profiles across the $DP$, $PP$, and GBS branches in module 1. \space
Let $t_{\mathrm{MB}}$ be the forward and backward time for one micro-batch
across all layers at the profiled TP and CP degrees, and
$G=b/(DP\cdot MBS)$ the number of micro-batches per optimizer step. Under a \space
balanced one-forward-one-backward (1F1B) schedule, a stage takes
$t_{\mathrm{MB}}/PP$ per micro-batch. Filling and draining the pipeline adds
$PP-1$ slots, giving
$\widehat{T}_{\mathrm{comp}}(x)=(G+PP-1)t_{\mathrm{MB}}/PP$. Combining it with
pipeline communication, data-parallel communication, and optimizer time gives
\begin{equation}
\small
    \widehat{f}_{\mathrm{thr}}(x)
    =
    \frac{b\,N_{\mathrm{tok}}}
    {\widehat{T}_{\mathrm{comp}}(x)
     +\widehat{T}_{\mathrm{pp}}(x)
     +\widehat{T}_{\mathrm{dp}}(x)
     +\widehat{t}_{\mathrm{opt}}(x)},
    \label{eq:throughput-estimation}
\end{equation}
where $N_{\mathrm{tok}}$ is the number of tokens per sequence,
$\widehat{T}_{\mathrm{pp}}$ and $\widehat{T}_{\mathrm{dp}}$ are pipeline- and
data-parallel communication times, and $\widehat{t}_{\mathrm{opt}}$ is optimizer
time. Recent executions calibrate these terms and the throughput variance.
For each GPU count, LIBRA times complete optimizer steps for a layout with
the largest feasible DP degree. These steps include gradient accumulation,
communication, and the optimizer update. For a fixed layout, GBS
sets the number of accumulated micro-batches $G$. This lets one step-time
measurement support several GBS estimates.

\textit{Throughput-aware Pruning.} \space
Module 2 ranks layouts at the same GPU count and GBS using
Eq.~\ref{eq:throughput-estimation}. Each box reports predicted throughput in K
tok/s as the mean $\pm$ one standard deviation. LIBRA retains the best
configuration and configurations whose intervals overlap the best interval. It
prunes configurations whose upper bounds are lower, leaving
throughput-competitive configurations for monitoring.
LIBRA explores a higher TP, CP, or PP degree when it allows a larger MBS.
If the largest feasible MBS stays the same, it stops expanding that degree.
When MBS grows, LIBRA compares mean throughput with the best layout already
found at each common GBS on the same GPU count. The search continues in
that direction if the new layout wins at least one comparison. The resulting
table gives AGPE a layout for each feasible pair of GPU count and GBS.

\textbold{Online Iterative Pruning via Convergence Monitoring.}
After Level 1, LIBRA maintains the remaining GBS values as an ordered pool.

\textit{Convergence Monitoring.}
Module 3 starts \allowbreak\space
one branch for each \space selected GBS from the same checkpoint. The checkpoint contains the same model,
optimizer, scheduler, and consumed-token state.
A window covers the incumbent and neighboring larger batches. Each branch
uses the layout selected for its GPU group. Evaluations run for a short
token horizon or fixed loss decrement. Fig.~\ref{fig:throughput-extrapolation}
shows the token-horizon case, where all branches process the same number of
tokens and record loss before and after the window. These measurements update
Eq.~\ref{eq:convergence-progress}'s posterior.

The monitoring window exposes differences in loss progress but is shorter than
a training phase. For example, a LLaMA-2 13B candidate with sequence length
4,096 and GBS 8 processes 3.3--6.6M tokens per window. Using the same training
state and stopping rule makes progress per token comparable across GBS and
parallel configurations.

\textit{CBS-aware GBS Pruning.}
Module 4 compares the Level 1 throughput posterior and module 3 convergence
posterior across the ordered GBS pool. The CBS insight is that progress per
token diminishes at large GBS while the useful batch regime moves upward.
LIBRA uses this insight for boundary pruning without estimating CBS, treating
the largest active GBS as the boundary. The joint confidence test
prunes a GBS only when its convergence upper bound is below the nearest
retained GBS's lower bound and its throughput upper bound offers less than 3\%
gain over it. Pareto-competitive or uncertain GBS values remain. This condition
must hold for two consecutive monitoring windows before stopping a boundary
candidate or upward expansion.
A smaller GBS that loses to a larger neighbor in time to the next milestone
can leave the pool. A larger neighbor with higher throughput remains eligible
at a later checkpoint, when its progress per token may improve relative to
the incumbent. The full GPU allocation sets an upper limit on this search.
Once doubling GBS adds only a marginal throughput gain over the best layout at
the preceding GBS, that preceding GBS is the last one to monitor.

In Fig.~\ref{fig:throughput-extrapolation}, module 3 shows three monitoring
branches for readability, while module 4 also uses the same-checkpoint
observation for GBS 32. GBS 32 is pruned, GBS 64 and 128 are retained, and GBS
256 is the boundary. \space
A competitive boundary below this limit admits the next feasible GBS. Retained neighbors can be
reconsidered without reopening the full space. Repeating this process produces
batch-size ramp-up without a predefined schedule.
Once the controller has compared the uppermost batch or the CBS test stops
expansion, fixed training uses the selected GBS and its profiled layout on
the full GPU allocation. The deepest compatible
checkpoint lets the selected configuration resume without repeating the
monitoring window.

\subsection{Acquisition-Guided Parallel Evaluation (AGPE)}
\label{sec:parallel-evaluation}
The Bayesian co-tuner in Section~\ref{sec:bayesian-cotuning} uses qNEHVI to
produce a candidate set $\mathcal{Q}_t$ and marginal acquisition values
$a_t(x)$, deciding \emph{what} to evaluate but not \emph{how} to map evaluations
to GPUs. For each ready candidate $x$, $w(x)$ is its GPU demand and
$\widehat{d}_t(x)$ is its estimated evaluation time. Their product
$\widehat{c}_t(x)=w(x)\widehat{d}_t(x)$ is the estimated GPU-time cost. AGPE
uses $p_t(x)=a_t(x)/\widehat{c}_t(x)$ as the scheduling priority. Each
scheduling event selects a feasible subset by solving
\begin{equation}
\small
\begin{aligned}
    \max_{z_x\in\{0,1\}} \quad
        & \sum_{x\in\mathcal{Q}_t} p_t(x)z_x \\
    \mathrm{s.t.}\quad
        & \sum_{x\in\mathcal{Q}_t} w(x)z_x \le W_{\mathrm{free}},
\end{aligned}
\label{eq:evaluation-packing}
\end{equation}
where $z_x$ is a binary admission variable, $\mathcal{Q}_t$ is the ready queue,
and $W_{\mathrm{free}}$ is the number of free GPUs. Topology and divisibility
restrict placement. Equation~\ref{eq:evaluation-packing} uses qNEHVI's
acquisition value while accounting for the expected GPU-time spent obtaining
the observation.

AGPE estimates throughput evaluation time from profile steps and recent
step-time measurements. For a fixed token horizon, it estimates convergence
evaluation time by dividing the horizon by the Level 1 throughput estimate.
Recent monitoring windows update estimates for evaluations that stop at a fixed
loss decrement. After admission, AGPE compares feasible placements using
$w(x)$ and $\widehat{d}_t(x)$ and chooses a placement that balances predicted completion times. Completed evaluations update time estimates.

\begin{figure}[t]
    \centering
    \begin{subfigure}[t]{\columnwidth}
        \centering
        \includegraphics[width=.9\linewidth]{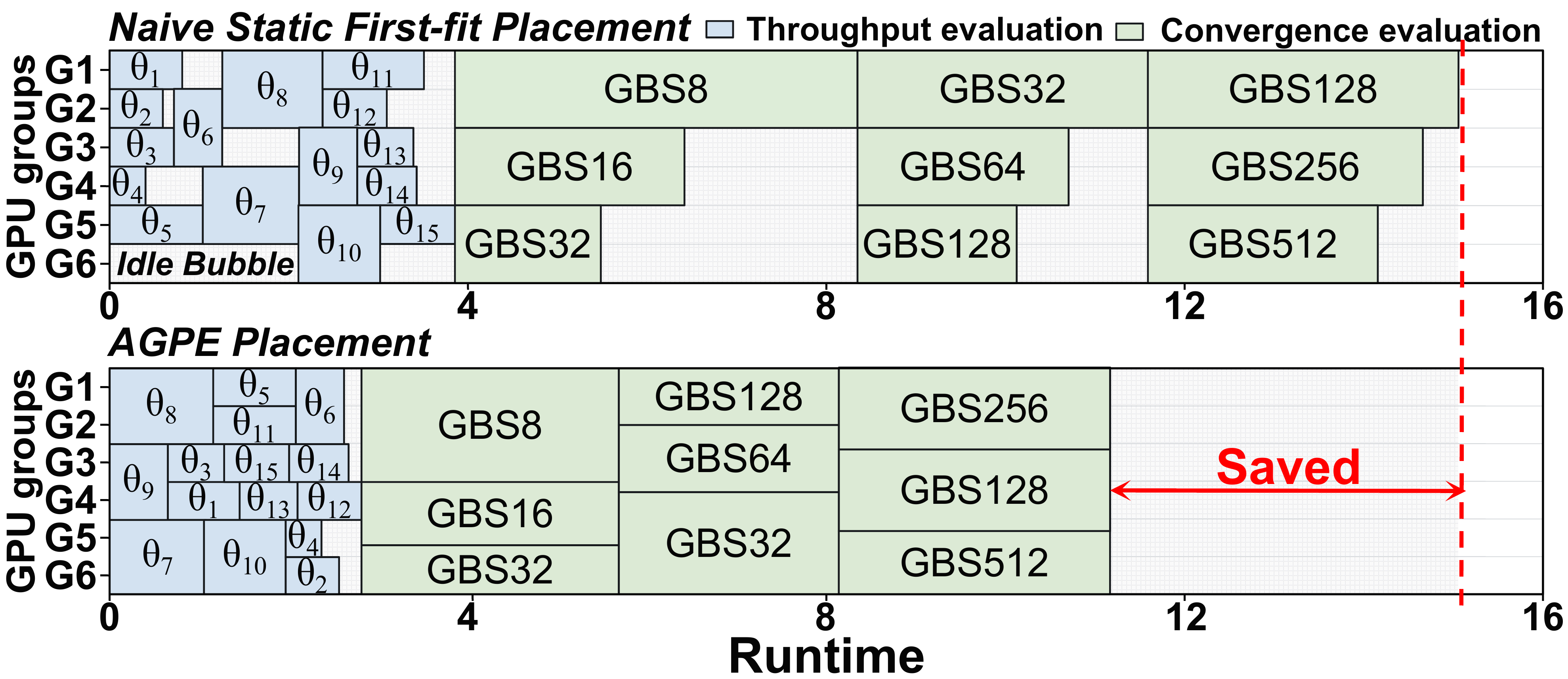}
        \caption{Evaluation placement.}
    \end{subfigure}
    \begin{subfigure}[t]{\columnwidth}
        \centering
        \includegraphics[width=.9\linewidth]{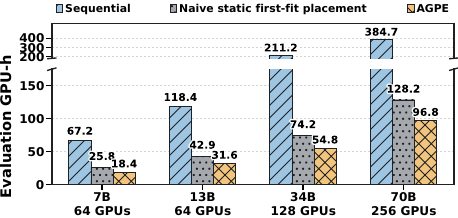}
        \caption{Evaluation cost.}
    \end{subfigure}
    \caption{Evaluation placement and cost. (a) Schematic showing how AGPE reduces idle GPU bubbles relative to naive static first-fit placement. (b) Total GPU-hour cost of sequential evaluation, naive static first-fit placement, and AGPE.}
    \label{fig:agpe-efficiency}
    \Description{A GPU-group timeline compares naive static first-fit placement
    with AGPE, and a bar chart compares the GPU-hour cost of sequential
    evaluation, naive static first-fit placement, and AGPE across four workloads.}
\end{figure}

Algorithm~\ref{alg:agpe} converts acquisition-ranked proposals into
asynchronous execution.
\begin{algorithm}[t]
\footnotesize
\DontPrintSemicolon
\caption{Our proposed AGPE method.}
\label{alg:agpe}
\KwIn{co-tuner $\mathcal{B}_t$, candidate pool
$\widetilde{\mathcal{X}}_t$, data $\mathcal{D}_t$, checkpoint $c_t$,
GPU groups $\mathcal{G}$}
$\mathcal{R}\leftarrow\emptyset$ \tcp*{running evaluations}
\While{\textsc{TuningActive}$(s_t)$}{
    $\mathcal{P}\leftarrow\textsc{Candidates}(\mathcal{R})$
    \tcp*{pending points}
    $(\mathcal{Q},\mathbf{a})\leftarrow
    \textsc{Propose}(\mathcal{B}_t,\widetilde{\mathcal{X}}_t,
    \mathcal{D}_t,\mathcal{P})$\;
    $(\mathbf{w},\widehat{\mathbf{d}})\leftarrow
    \textsc{EstimateResources}(\mathcal{Q},\mathcal{D}_t)$\;
    $\mathcal{A}\leftarrow
    \textsc{AcquisitionPack}(\mathcal{Q},\mathbf{a},\mathbf{w},
    \widehat{\mathbf{d}},\textsc{Free}(\mathcal{G}))$\;
    \ForEach{$(x,g)\in\mathcal{A}$}{
        $\mathcal{R}\leftarrow\mathcal{R}\cup
        \{\textsc{Launch}(x,g,c_t)\}$\;
    }
    \If{$\mathcal{R}=\emptyset$}{\textbf{break}\;}
    $e\leftarrow\textsc{WaitNextEvent}(\mathcal{R})$
    \tcp*{monitoring or terminal event}
    $(x,\mathbf{y},\Sigma,\mathrm{status})\leftarrow
    \textsc{Observe}(e)$\;
    \If{$\mathrm{status}\notin\{\mathrm{failed},\mathrm{timeout}\}$}{
        $\mathcal{D}_t\leftarrow\mathcal{D}_t
        \cup\{(x,\mathbf{y},\Sigma)\}$\;
        $\mathcal{B}_t\leftarrow
        \textsc{Update}(\mathcal{B}_t,\mathcal{D}_t)$\;
        $\widetilde{\mathcal{X}}_t\leftarrow
        \textsc{KeepPruneExpand}(\widetilde{\mathcal{X}}_t,x,\mathcal{D}_t)$\;
        \If{$\mathrm{status}=\mathrm{running}\wedge$
        \textsc{ConvergenceEval}$(x)\wedge
        \textsc{CBSStop}(x,\mathcal{D}_t)$}{
            \textsc{Terminate}$(e)$; $\mathrm{status}\leftarrow\mathrm{stopped}$\;
        }
    }
    \If{$\mathrm{status}\in\{\mathrm{complete},\mathrm{stopped},
    \mathrm{failed},\mathrm{timeout}\}$}{
        $\mathcal{R}\leftarrow\mathcal{R}\setminus\{e\}$;
        \textsc{Release}$(e,\mathcal{G})$\;
    }
}
\Return{$\mathcal{B}_t,\widetilde{\mathcal{X}}_t,\mathcal{D}_t$}\;
\end{algorithm}
At each scheduling event, AGPE reports running evaluations as pending to qNEHVI,
estimates the GPU demand and evaluation time of each proposal, and packs the
highest-priority proposals into the free GPU groups (Lines~3--8).
Fig.~\ref{fig:agpe-efficiency}(a) gives a schematic comparison with naive
static first-fit placement, which places evaluations in arrival order without
using estimated evaluation time. It illustrates how AGPE uses GPU demand and
estimated evaluation time to reduce idle GPU bubbles, while the exact placement
is determined at runtime.

AGPE waits for the next monitoring or terminal event and records its noisy
observation (Lines~11--20). A monitoring event updates the posterior and
candidate pool while the evaluation remains active. A terminal event releases
its resources. The next iteration uses the new data to recompute evaluation-time
estimates (Line~5) before scheduling again, without waiting for a fixed-width
round.

After each window, the CBS-aware joint confidence test retains, prunes, or expands
the ordered GBS pool and applies early stopping (Lines~16--20). \space
For example, when two GPU groups are free, AGPE can run the incumbent GBS
and its next larger neighbor concurrently. The throughput of each group
determines how long its evaluation occupies those GPUs. To compare the two
batches for training, LIBRA combines each branch's measured loss reduction
per token with the throughput of its proposed deployment layout in
Eq.~\ref{eq:deployment-time}. Groups become available for new evaluations
as their windows finish.

Fig.~\ref{fig:agpe-efficiency}(b) includes throughput and convergence
evaluations. AGPE uses 18.4--96.8 GPU-hours, compared with 25.8--128.2
GPU-hours for naive static first-fit placement and 67.2--384.7 GPU-hours for
sequential evaluation.
The sequential baseline completes every evaluation without pruning or
early stopping. Naive static first-fit placement uses the same updates
as AGPE but launches candidates in queue order. It can start candidates before
an observation removes them from the queue. AGPE returns
high-priority observations earlier, removes candidates before launch,
and stops weak evaluations. Estimated-time placement explains the shorter
completion time in Fig.~\ref{fig:agpe-efficiency}(a), while fewer
evaluations and early stopping explain the GPU-hour reduction in
Fig.~\ref{fig:agpe-efficiency}(b).

\begin{table}[t]
    \centering
    \caption{LIBRA's runtime interface.}
    \label{tab:system-interface}
    \footnotesize
    \setlength{\tabcolsep}{3.2pt}
    \renewcommand{\arraystretch}{1.04}
    \begin{tabular}{@{}p{0.25\columnwidth}p{0.67\columnwidth}@{}}
        \toprule
        \textbf{Role} & \textbf{Information} \\
        \midrule
        User inputs & Training task, GPU allocation, feasible parameter ranges,
        target validation loss \\
        Fixed recipe & Model, data order, optimizer, learning-rate schedule,
        sequence length \\
        Decisions & GBS and DP/TP/PP/CP/MBS configuration \\
        Observations & Throughput, monitored loss progress, resource demand,
        completion status \\
        Persisted state & Candidate queue, posterior data, incumbent, checkpoint
        metadata \\
        \bottomrule
    \end{tabular}
\end{table}
\section{System Implementation}
\label{sec:system-design}
This section describes how LIBRA turns optimization decisions into executable training jobs and incorporates their observations into the ongoing run.

\textbf{Runtime Control Workflow.}
LIBRA is an external control plane above training frameworks such as
Megatron-LM~\cite{shoeybi2019megatron,narayanan2021efficient}. Users provide the
training specification, cluster description, and GBS and parallelism
ranges, but no fixed configuration or ramp-up schedule. LIBRA validates
inputs and translates the decisions into executable jobs only at evaluation and reconfiguration boundaries.
Tab.~\ref{tab:system-interface} summarizes this interface and its runtime state.

\textbold{Candidate Construction.}
Each retained point $x=(b,\pi)$ becomes a record with GPU demand, evaluation
type, priority, checkpoint, and observations. Construction enforces
$\mathrm{DP}\!\times\!\mathrm{TP}\!\times\!\mathrm{PP}\!\times\!\mathrm{CP}=w_x$
for GPU-group size $w_x$, and GBS, sharding, and memory constraints.
Candidates sharing the $(TP,CP,MBS)$ tuple reuse one profile. \space
For a convergence evaluation, the controller looks up the selected layout
using the proposed GBS and group size. The same GBS can use a different
layout when deployed on all GPUs.

\textbold{Runtime Control and Reconfiguration.}
AGPE uses acquisition values, GPU demands, and estimated evaluation times to
assign candidates to disjoint GPU groups and launch them through Megatron-LM.
Completions asynchronously update the pruning, Bayesian, and time-estimation
modules before the next dispatch. Online candidates start from the same
checkpoint to keep convergence observations comparable. Once a candidate passes
the conservative rule in Eq.~\ref{eq:deployment-time}, the controller \space
switches at an optimizer-step boundary. It resumes from the deepest compatible
checkpoint on the selected branch, which includes the updates completed
during monitoring. A change in parallel layout redistributes model and
optimizer state to the new ranks. The data iterator and learning-rate
schedule continue from the saved training progress. Only the selected
branch's tokens advance the ongoing run. If the incumbent wins, training
continues from its own endpoint.

\begin{figure*}[t]
    \centering
    \includegraphics[width=0.9\textwidth]{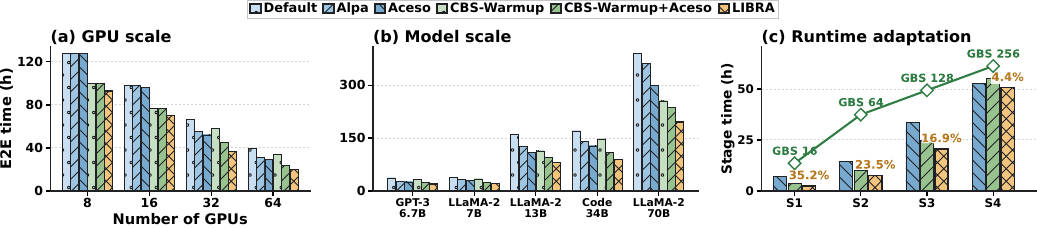}
    \caption{E2E time to the target validation loss and runtime
    adaptation. (a) LLaMA-2 7B on different GPU counts. (b) Different
    models on 64/64/64/128/256 GPUs. (c) Stage time and how GBS changes across
    different stages.}
    \label{fig:e2e-results}
    \Description{Three plots compare E2E time across GPU and model scales
    and show per-stage runtime adaptation for LLaMA-2 13B.}

\end{figure*}

\textbf{Other Details.} LIBRA comprises roughly 7K lines of Python above
Megatron-LM. It draws on backend timers and communication statistics and
launches retained configurations through the Megatron-LM entry point, leaving
model operators, collectives, pipeline schedules, the optimizer, and the
learning-rate schedule untouched, and persists its search state alongside checkpoints. Failed or timed-out evaluations release their GPUs without
updating the posterior, whereas early-stopped ones retain their observations; if
no candidate passes the conservative switching rule, training proceeds with the
current configuration.

\section{Evaluation}
We evaluate LIBRA across model and GPU scales, end to end and by component effectiveness and overhead.

\subsection{Experimental Setup}
\label{sec:eval_setup}

\textbf{Hardware and Software.}
Experiments use 1--32 nodes of a Hygon AMD GPU cluster, totaling
8--256 GPUs. Each node has two 64-core Hygon C86 CPUs and eight Hygon AMD GPUs
with 64~GB memory, based on the \texttt{gfx936} architecture with 80 compute units. GPUs connect
through PCIe Gen5 x16 and a fully connected Hygon hyswitch fabric. Each node
has four 400~Gbps InfiniBand NICs. The stack comprises Anolis OS 8.9, a ROCm
6.3-compatible GPU runtime, RCCL 2.22.3, PyTorch 2.4.1, and Megatron-LM
0.12.0. All experiments run on this Hygon AMD GPU platform, while the
rest of the paper uses GPU as the accelerator term. LIBRA is not tied to this
hardware. Runtime profiling recalibrates its performance estimates on the
deployment platform rather than relying on a fixed hardware model.

\textbf{Models and Dataset.}
We train GPT-3 6.7B, LLaMA-2 7B/13B/70B, and Code LLaMA 34B on OpenWebText.
Sequence lengths are 1,024 for GPT-3, 4,096 for LLaMA-2, and 4,096 for Code
LLaMA. Target validation losses are 2.45 for GPT-3, 2.37, 2.26, and 2.22 for
LLaMA-2 7B, 13B, and 70B, and 2.25 for Code LLaMA.

\textbf{Baselines.} Baselines span expert practice, parallel tuning, batch
adaptation, and combined tuning. All share the candidate space and
Megatron-LM~\cite{shoeybi2019megatron} backend. (i) \textbf{Default} uses a
per-model batch policy and one manual parallel configuration.
(ii) \textbf{Alpa}~\cite{zheng2022alpa} combines that policy with
inter- and intra-operator parallel search. (iii) \textbf{Aceso}~\cite{aceso}
uses bottleneck alleviation for throughput-oriented parallel search.
(iv) \textbf{CBS-Warmup}, based on Merrill et al.~\cite{merrill2025critical},
doubles GBS when CBS exceeds twice the batch while fixing parallel
configuration. (v) \textbf{CBS-Warmup+Aceso} applies Aceso's throughput-best
configuration at each GBS.

The composition tests whether connecting batch and parallel tuners explains
LIBRA's gain. All methods share initialization, data order, optimizer, recipe,
allocation, constraints, and target loss. E2E time includes online profiling,
evaluation, and reconfiguration, but excludes offline schedule and configuration
construction, favoring the baselines.

For the HPO comparison in Section~\ref{sec:eval_e2e}, we adapt
ASHA~\cite{li2018massively} and BOHB~\cite{falkner2018bohb} to the same joint
domain after offline pruning. Each candidate pairs one GBS with one parallel
configuration. Monitoring tokens define fidelity, and progress per unit time is the
promotion signal. To isolate search policy, all methods use the same GPU-group
placement, asynchronous feedback, and fidelity limits.

\textbold{Training Settings.}
All runs start from the same model state and train from the beginning. Methods
vary only GBS and parallel configuration within the same space, subject to
model, memory, divisibility, and allocation constraints. The optimizer,
learning-rate schedule, sequence length, data order, validation cadence, target
loss, and checkpointed state remain identical. We report E2E time from job
launch until validation first reaches the target, including online profiling,
convergence monitoring, candidate scheduling, checkpoint I/O, and
reconfiguration. Throughput excludes warm-up steps and uses the training loop's
token accounting.

\subsection{E2E Time-to-Convergence}
\label{sec:eval_e2e}
Fig.~\ref{fig:e2e-results}(a) and (b) cover 8--64 GPUs and models from GPT-3
6.7B to LLaMA-2 70B. Across the distinct settings, LIBRA's reductions relative
to Alpa and Aceso average 31.1\%. Relative to Alpa, reductions grow from
27.5\% to 36.0\% over 8--64 GPUs and from 27.6\% to 46.0\% across model scales.
Relative to Aceso, model-scale reductions range from 21.2\% to 34.8\%. For
GPT-3 6.7B, LIBRA reaches the target in 20.02 hours using 14.12B tokens, versus
Alpa's 27.65 hours and 24.09B tokens, despite lower throughput (196K versus
242K tokens/s). Its advantage over the best alternative grows from 7.1--8.7\%
at 8--16 GPUs to 15.6--17.4\% at 32--64 GPUs, and ranges from 13.4--18.6\%
across model scales for which all baselines are available. Thus, neither parallel tuning nor independent
batch adaptation explains the gain. Fig.~\ref{fig:e2e-results}(c) examines a
13B run. Each stage is a validation-progress interval delimited
by LIBRA's reconfiguration points. Bars compare the time each method needs to
traverse the same interval, while the line traces LIBRA's GBS. LIBRA reduces
stage time over the best alternative by 35.2\%, 23.5\%, 16.9\%, and 4.4\% as
GBS increases from 16 to 256. Its largest gains therefore arise before the final
large-batch regime.

\begin{figure}[t]
    \centering
    \includegraphics[width=.8\columnwidth]{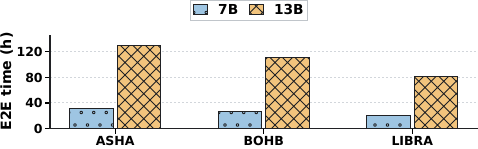}
    \caption{Comparison of HPO baselines and LIBRA.}
    \label{fig:hpo-comparison}
    \Description{Grouped bars compare the E2E time of ASHA, BOHB, and
    LIBRA on LLaMA-2 7B and 13B.}
\end{figure}

Fig.~\ref{fig:hpo-comparison} compares HPO baselines and LIBRA in the same
joint domain. ASHA, BOHB, and LIBRA reach the target in
31.83/27.37/20.17 hours on 7B and 129.47/110.62/81.37 hours on 13B,
respectively. BOHB improves on ASHA through model-based proposals, but both use
a scalar promotion signal across a broader candidate set. LIBRA combines
two-objective acquisition with hierarchical pruning and therefore reaches the
target sooner even when the HPO baselines retain its parallel execution layer.

\begin{figure}[t]
    \centering
    \includegraphics[width=\columnwidth]
    {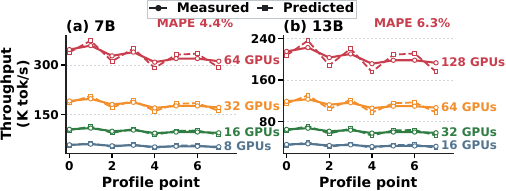}
    \caption{Comparison of predicted throughput in LIBRA's \textit{performance
    estimation} (Sec.~\ref{sec:hierarchical-pruning}) with measured
    throughput.}
    \label{fig:performance-estimation}
    \Description{Measured and predicted throughput curves across representative profiling points for LLaMA-2 7B and 13B.}
\end{figure}

\subsection{Accuracy of Performance Estimation}
Fig.~\ref{fig:performance-estimation} compares measured and predicted
scale-out throughput.
The predicted curves follow the measured scale-out trends, with mean absolute
percentage error (MAPE) of 4.4\% for 7B over 8--64 GPUs and 6.3\% for 13B over
16--128 GPUs. This accuracy lets offline pruning rank unmeasured batch-level
configurations while retaining configurations whose uncertainty overlaps the best estimate.
The error remains low as GPU count increases, showing that the extrapolation
preserves its accuracy across the evaluated scale-out range.

Tab.~\ref{tab:search-reduction} follows candidates through both pruning levels.
Feasibility checks reduce 1.52--2.44M full tuples to 11.3--24.6K feasible
candidates. Profile reuse and performance extrapolation leave 72--96
throughput-competitive configurations after O1, and convergence monitoring leaves 6--8
after O2. LIBRA measures only 68--98 profiles in 7--9 parallel rounds.
Hierarchical pruning, profile reuse, AGPE, and early stopping together save
72.6--74.8\% of evaluation GPU-hours over sequential evaluation.
The reductions are cumulative. Feasibility checks remove invalid configurations,
offline pruning avoids most configuration measurements, and online pruning reduces the
retained joint domain to 6--8 candidates. The stable savings across workloads
show that this hierarchy controls evaluation cost as model and GPU scales grow.

\begin{table}[t]
    \centering
    \caption{Search-space size and evaluation cost after hierarchical pruning and AGPE. Full, Feasible, O1, and O2 denote the unfiltered space and the spaces after the feasibility check, offline pruning, and online pruning. Cost is in GPU-hours; Reduction is relative to sequential evaluation.}
    \label{tab:search-reduction}
    \footnotesize
    \setlength{\tabcolsep}{2.0pt}
    \renewcommand{\arraystretch}{1.08}
    \begin{tabular}{@{}cc
        >{\columncolor{evalbody}}c
        >{\columncolor{evalbody}}c
        >{\columncolor{evalbody}}c@{}}
        \toprule
        \multirow{2}{*}{\makecell{\textbf{Model}\\\textbf{size}}} &
        \multirow{2}{*}{\makecell{\textbf{Search-space size}\\
        \textbf{(Full/Feasible/O1/O2)}}} &
        \multicolumn{3}{>{\columncolor{evalhead}}c@{}}
        {\textbf{Evaluation cost}} \\
        & &
        \cellcolor{evalhead}\textbf{Sequential} &
        \cellcolor{evalhead}\textbf{LIBRA} &
        \cellcolor{evalhead}\textbf{Reduction} \\
        \midrule
        7B  & 1.52M/19,872/96/7 & 67.2  & 18.4 & 72.6\% \\
        13B & 1.52M/11,264/72/6 & 118.4 & 31.6 & 73.3\% \\
        34B & 1.91M/15,360/88/7 & 211.2 & 54.8 & 74.1\% \\
        70B & 2.44M/24,576/96/8 & 384.7 & 96.8 & 74.8\% \\
        \bottomrule
    \end{tabular}

\end{table}

\subsection{Overhead of LIBRA}

\begin{figure}[tbp]
    \centering
    \includegraphics[width=.95\columnwidth]{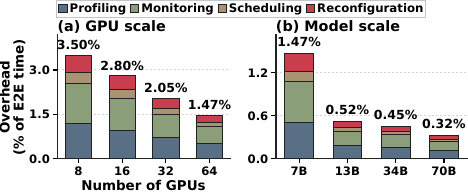}
    \caption{Runtime overhead of LIBRA, measured as additional critical-path
    time divided by E2E training time across different models and GPU counts.
    The bars show profiling, monitoring, scheduling, and reconfiguration.}
    \label{fig:runtime-overhead}
    \Description{Breakdown of LIBRA runtime overhead across GPU counts and model sizes.}
\end{figure}

Fig.~\ref{fig:runtime-overhead} reports the additional critical-path time from
profiling, monitoring, scheduling, and reconfiguration as a percentage of E2E
training time. The controller timestamps intervals in which a module delays the
training or reconfiguration action, excluding work that overlaps execution on
other GPU groups. This metric differs from the evaluation GPU-hours in
Fig.~\ref{fig:agpe-efficiency}(b) and Tab.~\ref{tab:search-reduction}, which
measure GPU use rather than the time remaining on the critical path after
concurrent evaluations overlap. For 7B, overhead falls from 3.5\% at 8 GPUs to 1.47\% at 64 GPUs.
Across the model-scale settings, it is 1.47/0.52/0.45/0.32\% for
7B/13B/34B/70B, respectively. Bounded windows and concurrent placement keep cost low, and all
E2E results include it. The declining fraction at larger scales indicates that
LIBRA's runtime control does not become a scalability bottleneck.

In addition, LIBRA's memory footprint is confined to host-side search and
checkpoint metadata and remains under 51.8~MiB per run because hierarchical
pruning bounds the retained candidate set. Evaluations reuse Megatron-LM's
allocator and add no persistent device-memory allocation.

\begin{figure}[tbp]
    \centering
    \includegraphics[width=.95\columnwidth]{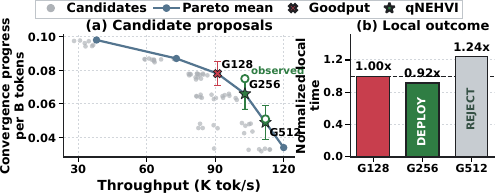}
    \caption{Comparison of two multi-objective optimization methods: qNEHVI and
    Goodput. (a) The configurations selected by both methods lie on the Pareto
    frontier. (b) Normalized local time for the observed candidates at the
    checkpoint.}
    \label{fig:qnehvi-selection}
    \Description{A Pareto plot and normalized local-time bars show that short
    monitoring selects GBS 256 over the GBS 128 Goodput choice and rejects GBS 512.}
\end{figure}

\subsection{Ablation Study}
\label{sec:eval_ablation}

\begin{figure}[tbp]
    \centering
    \includegraphics[width=.95\columnwidth]{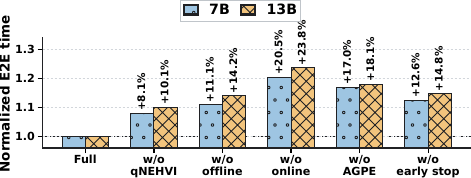}
    \caption{Ablation study for LLaMA-2 7B and 13B.}
    \label{fig:removal-ablation}
    \Description{Grouped bars show normalized E2E time after removing qNEHVI, either pruning level, AGPE, or early stopping.}
\end{figure}

\textbf{Multi-objective Acquisition.}
Figure~\ref{fig:qnehvi-selection} compares candidate selection by goodput and
qNEHVI at a 13B checkpoint. Goodput selects GBS 128, whereas qNEHVI also
proposes GBS 256 and 512. This difference comes from how the methods use
throughput and convergence. Goodput combines them into a single score through
statistical efficiency, fixing their trade-off in advance. In contrast, qNEHVI
keeps the two objectives separate and accounts for uncertainty when selecting
candidates. Short monitoring shows that GBS 256 and 512 require
$0.92\times$ and $1.24\times$ the local time of GBS 128, respectively
(Eq.~\ref{eq:deployment-time}, normalized to GBS 128). The controller therefore
deploys GBS 256 and rejects GBS 512. Although these differences are modest,
keeping the two objectives separate distinguishes a beneficial candidate from
a slower one that goodput cannot identify. Replacing qNEHVI with goodput
increases E2E time by 8.1\% and 10.1\% for 7B and 13B, respectively
(Figure~\ref{fig:removal-ablation}).

\textbf{Component Removal.}
Fig.~\ref{fig:removal-ablation} \space compares the full system with five variants, each with one mechanism removed. ``w/o qNEHVI'' replaces qNEHVI with goodput, ``w/o offline'' removes performance pruning, ``w/o online'' removes CBS-aware pruning, ``w/o AGPE'' evaluates candidates sequentially, and ``w/o early stop'' completes every monitoring window. These add 8.1--10.1\%, 11.1--14.2\%, 20.5--23.8\%, 17.0--18.1\%, and 12.6--14.8\% to E2E time, respectively.

CBS-aware online pruning \space has the largest effect. Without it, a GBS that makes little progress per token can be tested again at later checkpoints. Its fastest layout can pass offline pruning because that comparison uses throughput at a fixed GBS. Early stopping releases GPUs from a weak branch within the current window. On both the 7B and 13B workloads, every removal increases E2E time by at least 8\%.

\section{Related Work}
\label{sec:related_work}

\textbf{Throughput-oriented Parallel Tuning.}
Distributed training systems automate planning across parallelization,
placement, and trade-offs involving memory and communication.
FlexFlow~\cite{flexflow} uses simulator-guided search over operator-level
parallelization. Alpa~\cite{zheng2022alpa} combines intra-operator sharding
with inter-operator pipelining. Aceso~\cite{aceso} searches multidimensional
parallel configurations by alleviating performance bottlenecks, while
AutoDDL~\cite{autoddl} and Hypertron~\cite{hypertron} use analytical models to
explore broader spaces. Together, these systems improve execution throughput
for fixed training computations. However, they do not use convergence feedback
to decide when GBS should change. LIBRA complements them by jointly adapting
batch size and parallel configuration because a throughput-optimal
configuration need not minimize time-to-convergence.

\textbf{Hyperparameter Optimization and Online Tuning.}
Multi-fidelity hyperparameter optimization reduces tuning cost by allocating
partial-training resources across configurations. Successive
Halving~\cite{jamieson2016nonstochastic}, Hyperband~\cite{li2017hyperband}, and
ASHA~\cite{li2018massively} terminate weak configurations early and promote
promising ones. BOHB~\cite{falkner2018bohb} and DEHB~\cite{awad2021dehb}
combine multi-fidelity allocation with model-based or evolutionary search,
while FABOLAS~\cite{klein2017fabolas}, BOCA~\cite{kandasamy2017boca}, and
trace-aware optimization~\cite{wu2020practical} use lower-fidelity observations
to guide selection. Ray Tune~\cite{liaw2018tune} provides distributed execution
for such methods. These methods evaluate candidates as separate
training runs. For LLM training, each batch-size candidate maps to multiple
parallel configurations, making HPO expensive. Lower-fidelity
observations reduce cost but do not remove this parallel configuration search.
LIBRA uses qNEHVI but structures optimization around hierarchical pruning and
concurrent evaluation within an ongoing run rather than launching full or
epoch-level runs.

\section{Conclusion and Future Work}
We presented LIBRA, a lightweight runtime framework that jointly adapts batch size and parallel configuration to minimize time-to-convergence. LIBRA combines qNEHVI-based multi-objective co-tuning, hierarchical pruning, and AGPE to make runtime candidate search and evaluation practical. Across a range of LLMs and cluster scales of up to 256 GPUs, LIBRA reduces E2E time-to-convergence by 31.1\% on average compared with state-of-the-art methods, while its runtime tuning accounts for only 0.3--3.5\% of E2E training time. These results demonstrate the benefit of optimizing throughput and convergence jointly rather than in isolation.

Several directions remain open. Our prototype is built on Megatron-LM, and porting the controller to other training frameworks would test how far the design generalizes across runtime interfaces.
Broader evaluation---covering additional model families, larger clusters, and mixture-of-experts architectures---would further characterize where joint co-tuning is most beneficial.

\bibliographystyle{ACM-Reference-Format}
\bibliography{refs}

\end{document}